\documentclass[12pt,preprint]{aastex}
\usepackage{graphicx}
\usepackage{amssymb}
\begin{document}

\title{A Deep Infrared Search for AXP 1E~1841$-$045.}
\author{Martin Durant}
\affil{Department of Astronomy and Astrophysics, University of
  Toronto,\\ 60 St. George St, Toronto, Canada}

\begin{abstract}
Multi-colour (JHK$_\textrm{S}$) imaging and photometry of the field of
the Anomalous
X-ray Pulsar AXP 1E~1841$-$045 is analysed in the light of new, accurate
coordinates from {\em Chandra} (Wachter et al, 2004). From excellent
quality images, we find multiple sources in and around the position
error circle. Of these, none can be confidently identified as the
infrared counterpart. The limiting magnitudes reached were $J=22.1$,
$H=20.7$ and $K_\textrm{S}=19.9$ (95\% confidence).
\end{abstract}
\maketitle

\section{Introduction}
The Anomalous X-ray Pulsars (AXPs) are a small group of young,
energetic neutron stars, whose luminosity is thought to be powered by
the decay of a super-strong magnetic field: {\em magnetars} (Thompson
\& Duncan, 1996). Since the
discovery of the first optical counterpart to an AXP (Hulleman, van
Kerkwijk \& Kulkarni, 2000), searches have been undertaken to identify
further optical and infrared counterparts in different colours. Due to
the large extinction to most of these sources, the infrared has proved
the more successful route. See Woods \& Thompson (2004) for a review
of the AXPs and their counterparts to date.

1E~1841$-$045 is located within the supernova remnant Kes~73, and has
a pulse period of 11.8s and a soft X-ray spectrum well-fitted by
either a black-body plus power-law or the sum of two black-bodies
with a fitted hydrogen absorption column of
$N_H=2.5\times10^{22}$cm$^{-1}$ (Morii et al., 2003; Gotthelf et al.,
2004). As a recent
surprise, this source was found by
Kuiper at al. (2004) to have hard, pulsed X-ray emission with a rising
power-law spectrum out to about 100keV. Since this, then, dominates
the emission energetics, it has prompted an ongoing revisal of
magnetar electrodynamics.

The supernova remnant Kes~73,
has an estimated age of $\sim$1500yrs (e.g. Gotthelf et al., 2002) and
using H~I
measurements towards the SNR, its distance has been determined as 6--7.5kpc (Sanbonmatsu
\& Helfand, 1992). Geometric alignment and the youth of
both the SNR and AXP (whose age is not easily determined, but is
of the order thousand of years) point to the association of the two being
real.

Mereghetti et al. (2001) were the first to search for an infrared/optical
counterpart to 1E~1841$-$045 by performing multi-colour imaging and
selected spectroscopy, but based on only {\em ROSAT} and {\em
  Einstein} positions.

Wachter et al. (2004) report a precise location for the AXP based on
new {\em Chandra} observations. They give the source's coordinates as
RA=18:41:19.343, Dec=-04:56:11.16, with a $3\sigma$ error radius of
$0.9''$. The images presented by these authors show an object in the
error circle that is either extended or made up of multiple sources.

Here we present deeper images taken in better seeing, from which we
attempt to identify the infrared counterpart to 1E~1841$-$045.

\section{Observation and analysis}
We imaged the field of 1E~1841-045 on the night of 5th June 2003 with
PANIC (Persson's Auxiliary Nasmyth
Infrared Camera, Martini et al., 2004), the 1k$\times$1k infrared imaging
array with $0.125''$ pixels on the Magellan Clay
Telescope\footnote{see {\tt
http://www.ociw.edu/lco/magellan/instruments/PANIC/panic/}}, under
excellent conditions. The total integration times were 1125s in J,
1825s in K$_\textrm{S}$, and 1825s in H, at seeing
of $\simeq0.35''$. A second J-band
integration was performed, but the seeing had
deteriorated, and so this is not included in the analysis.

Standard reduction was carried out to flat field and combine the frames using {\tt
IRAF}. The flat fields were derived by median combining many images
in each filter of a less crowded field. This proved more successful
than either screen flats or median images from the data frames
themselves. Photometry was
performed using {\tt DAOPHOT~II} (Stetson, 1987).

In order to calibrate the frames, short-exposure images were obtained
of standard stars, from Persson et al. (1998). Because of some light
cloud in patches on the
night in question, it was found more reliable to use standards from the
following night and find the magnitude transformation from one night
to the next using fields which were imaged on both nights (see Durant
\& van Kerkwijk, 2005). The offsets
were small in each band, $\simeq0.03$mag. The standards were taken at a
range of airmasses, so the zero point at the appropriate airmass could
be found (the variation with airmass is slight in the infrared in any
case). 

The magnitudes found for stars in the field tend to be fainter than
those found by Wachter et al. (2004) and Mereghetti et
al. (2001) by typically 0.3mag for their faintest stars (note that
there are also some substantial differences in the magnitudes presented by
these two sets of authors). This can be
attributed to the better seeing conditions,
which allowed sources to be separated which would otherwise have been 
blended in this extremely crowded field. Many of the stars
measured by Mereghetti et al. (2001) are saturated on our deeper
images, and so cannot be compared. We believe that the better
separation of sources entirely explains the discrepancy in measured
magnitudes.

An astrometric solution was found for the images based on 2MASS
sources (Curti et al., 2003) in the field. 86 stars were matched for J
in the $\simeq
2'\times 2'$ field, and after rejecting large residuals, the RMS
deviation in each coordinate was $\simeq0.1''$ with 71 points. The
error in connecting this image to the others is negligible in
comparison. The astrometric uncertainty is in connecting our images to the 2MASS
reference frame. Since Wachter et al.'s coordinates are also based on
2MASS stars, there should be no additional uncertainty in the astrometry.

\section{Results}

\begin{figure}
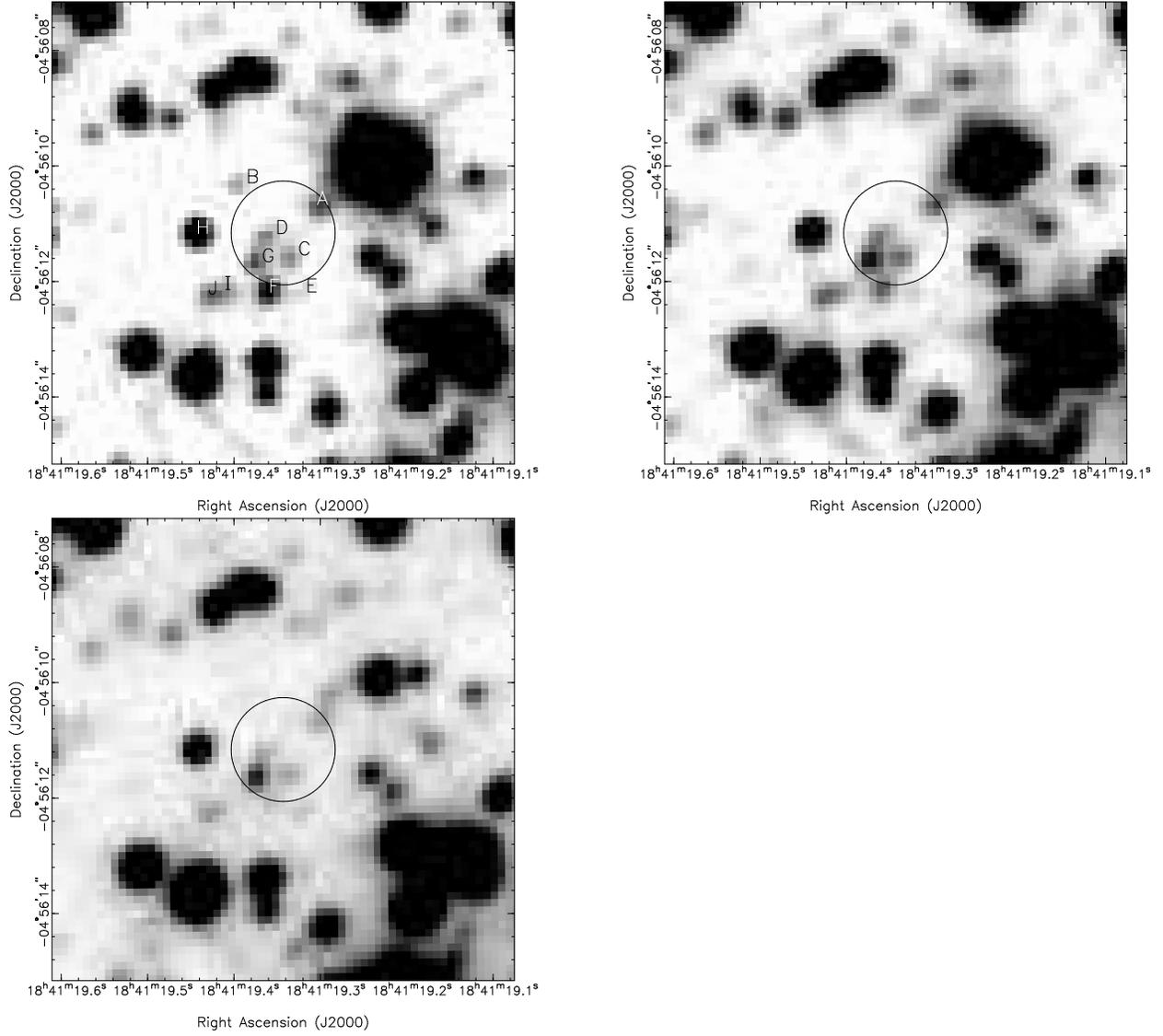

\begin{center}
\parbox{\hsize}{
\includegraphics[width=0.45\hsize,angle=270]{f1a.eps}
\includegraphics[width=0.45\hsize,angle=270]{f1b.eps}
\includegraphics[width=0.45\hsize,angle=270]{f1c.eps}}
\caption{Images of the field of 1E~1841-045 in the K$_\textrm{S}$-
  (top-left), H- (top right)
  and J-bands (bottom). In the left hand image are labelled
  the stars whose magnitudes are presented in Table \ref{mags}}\label{pics}
\end{center} 
\end{figure}

\begin{figure}[p]
\includegraphics[width=\hsize,angle=0]{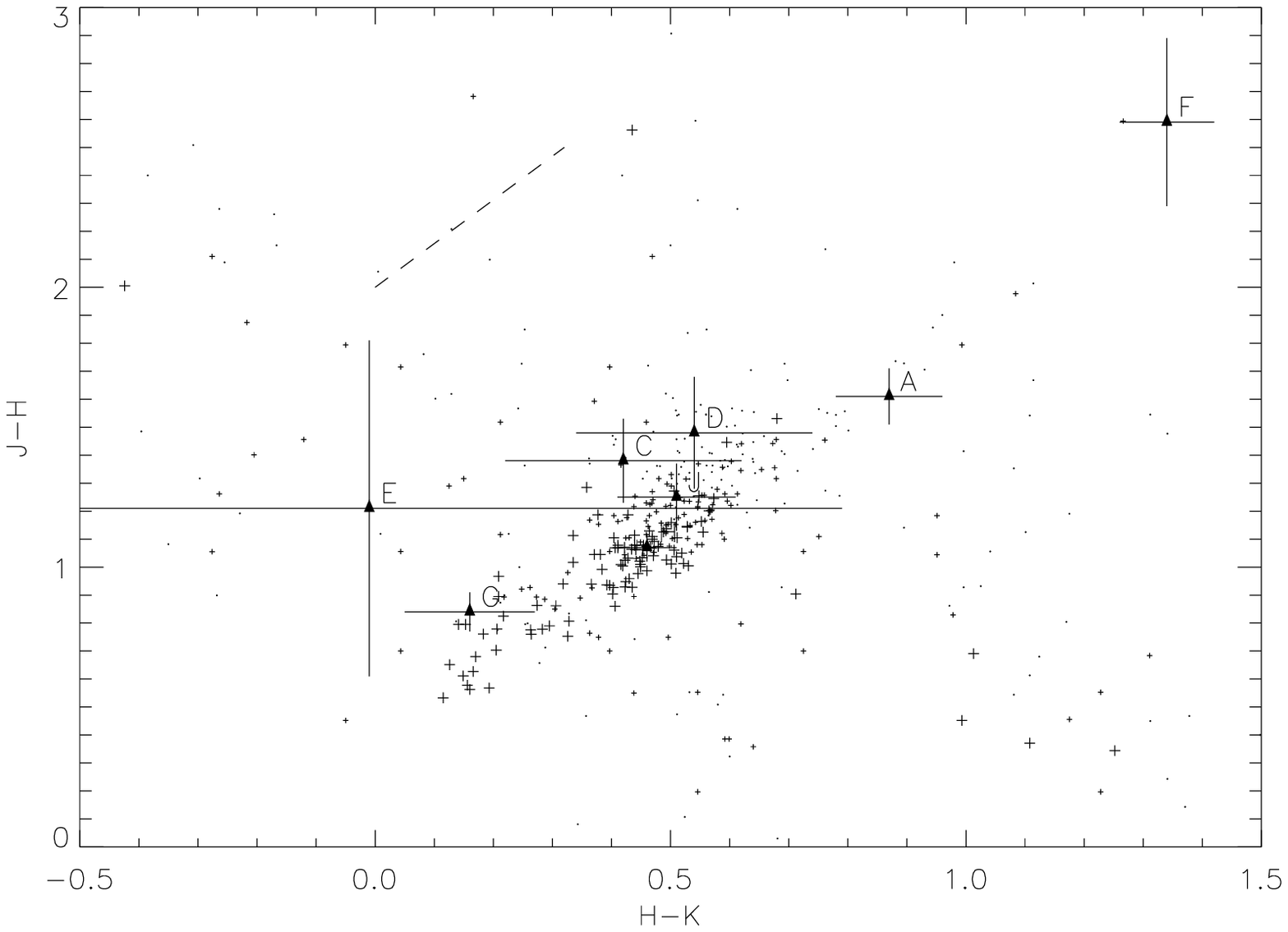}
\caption{Colour-colour diagramme of stars in the field of
  1E~1841$-$045. The stars in or close to the {\em Chandra}
  error circle are labelled with letters. Only stars with magnitude
  errors less than 0.15 in each band are plotted (comparable to stars
  C and D). Bigger symbols represent stars with lower magnitude errors
  in all three bands. The dashed line shows the effect of five
  magnitudes of visual extinction. AXPs are expected to lie below the
  bulk of the stars in this diagramme, based on 4U~0142+61 (Israel et
  al., 2004).}\label{cc} 
\end{figure}

Figure \ref{pics} shows the stacked images, with the position error
circle of radius $0.9''$ (3-$\sigma$ confidence) derived
by Wachter et al. (2004) overdrawn. Table \ref{mags} gives the magnitudes of
stars in and around the circle, as labelled on the images, and Figure
\ref{cc} shows those stars with three measured magnitudes on a
colour-colour diagramme compared to the rest of the stars in the
field. 

From Figure \ref{cc}, one sees that,
of the stars near the positional error circle (see Figure \ref{pics}),
none have significantly different colours to other stars in the
field. Note that the large scatter is due to the extreme crowding
in the field, particularly in K. This means that the measured
magnitude of a given star can be strongly affected by the halo of a
neighbouring brighter star. The magnitude limits reached at 95\%
confidence are: $J=22.1$, $H=20.7$ and $K_\textrm{S}=19.9$.

\begin{deluxetable}{clllll}
\tablecaption{Positions and magnitudes of stars within or near the {\em Chandra}
  error circle. \label{mags}}
\tablewidth{0pt}
\tablehead{\colhead{Star ID\tablenotemark{a}} & \colhead{R. A.} &
  \colhead{Dec} & \colhead{J} &
  \colhead{H} & \colhead{K$_\textrm{S}$} }
\startdata
A & 18:41:19.293 & -04:56:10.67 & 20.96(9) & 19.35(6) & 18.44(7) \\
B & 18:41:19.388 & -04:56:10.28 & $>22.1$  & 20.8(4) & 19.6(2) \\
C & 18:41:19.327 & -04:56:11.57 & 21.28(12) & 19.90(11) & 19.44(18) \\
D & 18:41:19.356 & -04:56:11.23 & 21.38(14) & 19.90(11) & 19.32(15) \\
E & 18:41:19.315 & -04:56:12.20 & 22.2(3) & 20.9(5) & 20.9(6) \\
F & 18:41:19.352 & -04:56:12.15 & 22.1(3) & 19.47(7) & 18.09(5) \\
G & 18:41:19.367 & -04:56:11.64 & 19.93(4) & 19.09(5) & 18.89(9) \\
H & 19:41:19.435 & -04:56:11.13 & 18.97(3) & 17.90(3) & 17.40(4) \\
I & 19:41:19.401 & -04:56:12.15 & $>22.1$ & 20.04(13) & 19.14(12) \\ 
J & 18:41:19.421 & -04:56:12.24 & 20.82(8) & 19.57(8) & 18.99(10)
\enddata
\tablecomments{~Magnitude limits are at 95\% confidence. The systematic
  uncertainty in position does not affect relative co-ordinates.}
\tablenotetext{a}{as labelled in Figure \ref{pics}. }
\end{deluxetable}

By fitting the X-ray spectrum with an absorbed black body plus
power-law spectrum, a value for the hydrogen column density can be
derived. Assuming the Predehl \& Schmitt (1995) relationship, this
translates to an extinction towards the source of $A_V\approx14$. With
the caveat that the intrinsic X-ray spectrum is not known, this
number provides an approximate measure of reddening. Figure
\ref{cc} shows that the effect of extinction means that one cannot
distinguish between an intrinsically hot but highly extincted source
and an intrinsically cool (i.e. red) source. Also note that, since
the main sequence is known to start around (0,0) on this diagramme,
the bluest sources here have $A_V\approx2$, although extinction is
known to increase rapidly in this direction (e.g. Drimmel et al.,
2003). 

Although an out-lier on Figure \ref{cc}, Star F is consistent with
being a very highly reddened red super-giant. De-reddening it with
$A_V=14$ would not place it below the bulk of the stars, as is the
case with 4U~0142+61 ($H-K_\textrm{S}=1.0(1)$, $J-H=1.2(2)$, Israel et
al., 2004, Hulleman et al., 2004). Whether these two objects
would be expected to have the same spectrum is an open question, as is
the appropriate value of reddening. Those AXPs with confirmed infrared
counterparts appear to have similar X-ray to infrared flux ratios
(Durant \& van Kerkwijk, 2005), and Star F would have both a much
brighter counterpart and much lower X-ray to infrared flux ratio than
4U~0142+61. Although stars with colours as red
as Star F are rare in the field, it cannot be presented as a likely
counterpart. It is worth mentioning that Star B, if close to the the magnitude
limit in J, would fall in the right region of Figure \ref{cc}, but
again this can hardly be more than a suggestion of a candidate
counterpart.

Comparing with the spectrum of 4U~0142+61 again (the brightest and best-measured
AXP, $J=22.3(1)$, $H=21.1(1)$,
$K_\textrm{S}=20.15(8)$), one would expect 1E1841$-$0145's magnitudes
to fall beyond the magnitude limits
given above, especially if the nominal reddening values to the two sources
are to be believed (which would make the magnitudes above fainter by about
2.5, 1.7 and 1.1 magnitudes respectively). Thus a non-detection here
does not imply that the
two spectra are necessarily different, but does demonstrate that this part of
the sky is so crowded that finding the counterpart will prove very
difficult. 

In conclusion, we have found the magnitudes of several sources in or
near the accurate {\em Chandra} error circle for the position of
1E~1841$-$045. Despite the depth and quality of the images, we find no
source which can be confidently presented as the likely
counterpart. Extremely deep images with narrow point-spread functions
will be required in order to find the counterpart to this AXP.

\end{document}